**Society's educational debts in biology, chemistry, and physics across race, gender, and class**
Ben Van Dusen[a]*, Jayson Nissen[b] and Odis Johnson[c]
[a]*School of Education, Iowa State University, Ames, USA;* [b]*Nissen Education Research and Design, Slidell, USA;* [c]*School of Education, John Hopkins University, Baltimore, USA*

bvd@iastate.edu
901 Stange Road,
Ames, IA
50011

**Abstract**
The success of collaborative instruction in helping students achieve higher grades in introductory science, technology, engineering, and mathematics (STEM) courses has led many educators and researchers to assume these methods also address inequities. However, little evidence tests this assumption. Structural inequities in our society have led to the chronic underrepresentation of Black, Hispanic, women, and first-generation students in STEM disciplines. Broadening participation from underrepresented groups in biology, chemistry, and physics would reduce social inequalities while harnessing diversity's economic impact on innovation and workforce expansion. We leveraged data on content knowledge from 18,791 students in 305 introductory courses using collaborative instruction at 45 institutions. We modeled student outcomes across the intersections of gender, race, ethnicity, and first-generation college status within and across science disciplines. Using these models, we examine the educational debts society owes college science students prior to instruction and whether instruction mitigates, perpetuates, or exacerbates those debts. The size of these educational debts and the extent to which courses added to or repaid these debts varied across disciplines. Across all three disciplines, society owed Black and Hispanic women and first-generation Black men the largest educational debts. Collaborative instructional strategies were not sufficient to repay society's educational debts.
    *Keywords*: QuantCrit, Equity, Science Education, Educational Debt, Intersectionality

**Society's educational debts in biology, chemistry, and physics across race, gender, and class**
**Introduction**

For more than a decade, broadening participation in the science, technology, engineering, and mathematics (STEM) disciplines has been considered critical for creating a more equitable and prosperous society (Committee on Prospering in the Global Economy of the 21st Century, 2007; 2010; Committee on Underrepresented Groups and the Expansion of the Science and Engineering Workforce Pipeline, 2011; Kozlowski *et al*., 2022). However, the culture of STEM disciplines continues to advantage higher socio-economic status (SES) White men over women and men from minoritized groups (Cech, 2022; Hatfield *et al*., 2022). In this investigation, we examine the inequalities in the STEM disciplines of biology, chemistry, and physics that extend from the intersection of racial, ethnic, gender, and social class subordination. Injustices within these science fields in higher education build on inequalities and inequities in opportunities and funding in K-12 education (Cheryan *et al*., 2017). Ladson-Billings (2006; 2007) characterized these educational inequalities as an educational debt that society has accrued over generations owed to students who are Black, Hispanic, Indigenous, and living in poverty. We use the term Hispanic while aware that the term Latinx has grown in use, primarily because the former reflects one's linguistic background instead of just geographic ancestry (MacDonald 2001), and language has additional significance in learning contexts.

To move from constructing students as deficient and as the source of the problem to shifting the onus onto those with the power to change the system, we apply her societal educational debts framework to investigate inequalities before and after instruction due to the intersections of sexism, racism, and classism. This analysis examines conceptual learning in introductory biology, chemistry, and physics courses that nearly all used student-centered, collaborative pedagogies. Introductory courses in these disciplines often act as barriers to getting STEM degrees (Matx *et al*., 2017; Hatfield *et al*., 2022). This has led to calls for new research into the roles that classroom culture and lectures play in negatively impacting student outcomes (National Academies, 2016; National Research Council, 2012).

We focused on student-centered, collaborative pedagogies because they are often perceived as capable of broadening the participation of groups traditionally underrepresented in STEM (Theobald *et al*., 2020). Student-centered, collaborative learning creates better student outcomes than lecture-based instruction (Theobald *et al*., 2020; Freeman *et al*., 2014; Hake, 1998; Van Dusen & Nissen, 2020a). It is not clear, however, if these improvements ameliorate, maintain, or exacerbate the inequities between men and women and White and Black, Hispanic, and Indigenous students (Johnson *et al*., 2020; Ernest *et al.* 2019; Rodriguez *et al*., 2012; Eddy & Hogan, 2014; Shortlidge *et al*., 2019; Wright *et al*., 2017). Very little quantitative research on these issues has taken an intersectional perspective that examines student outcomes among social identity groups. The analysis provides a baseline for understanding inequities in student-centered, collaborative science learning environments and identifies commonalities in inequities across disciplines and course features.

To support reader interpretation of this manuscript, we provide a list of statistical and equity-related terms in Table 1.

**Research Questions**

To better understand the intersecting roles that sexism, racism, and classism play in shaping learning for the students enrolled in biology, chemistry, and physics courses, we asked the following questions:
(1) How much educational debt in science knowledge does society owe introductory college science students due to racism, sexism, and classism prior to instruction?
(2) To what extent do introductory college science courses mitigate, perpetuate, or exacerbate society's educational debts?

(3) To what extent do the educational debts owed students vary across biology, chemistry, and physics?

Answers to these questions can inform how instructors and policy-makers can broaden participation across STEM disciplines and provide information for marginalized communities to advocate for effective and equitable STEM courses.

## Literature Review / Conceptual framework

There have been calls from across the science education research communities to examine inequities in student outcomes (e.g., National Academies of Sciences, Engineering, and Medicine, 2023; Committee on Underrepresented Groups and the Expansion of the Science and Engineering Workforce Pipeline, 2011; Physical Review Physics Education Research, 2020; Wilson-Kennedy *et al.*, 2022; Thompson *et al.*, 2020). Quantitative investigations of equity in physics (e.g., Traxler & Brewe, 2015; Van Dusen & Nissen, 2020a; 2020b; Nissen *et al.*, 2021; Rodriguez, 2012; Shafer *et al.*, 2021), chemistry (e.g., Dalgety & Coll, 2006; Lewis & Lewis, 2008; Rath *et al.*, 2012; Sunny *et al.*, 2017; Rocabado *et al.*, 2019; Bancroft *et al.*, 2019; Van Dusen & Nissen, 2022), and biology (e.g., Barr *et al.*, 2008; Eddy *et al.*, 2014; Wright *et al.*, 2016; Zuckerman *et al.*, 2021) have consistently found inequities in student learning, science identity, recruitment, and retention. Two areas of need in the literature are a more intersectional understandings how marginalized groups are currently being served in science courses and what interventions improve equity in student outcomes.

There is no consensus about what role active engagement instruction plays in the repayment or generation of educational debts. Active engagement instruction includes a broad range of pedagogical strategies that are grounded in the idea of students working collaboratively to construct their knowledge (Walshaw, 2004) and is often contrasted with traditional lecture-based pedagogy (Van Dusen & Nissen, 2020a). For example, Theobald *et al*. (2020) and Van Dusen & Nissen (2020a) found that high levels of active engagement were associated with more equality in student grades and course failure rates, while Johnson *et al*. (2020) and Ernest *et al*. (2019) found that active-learning classrooms were often associated with less equality in student participation in discussion which contributed to the marginalization of women.

Most of the science education research literature has not examined student outcomes from an intersectional perspective that accounts for the interaction effects between social identities (Kozlowski, 2022). Shafer *et al*. (2021) problematized the common practice of comparing non-Hispanic White and Asian physics students against all other racial and ethnic groups. In their analyses, the act of disaggregating across racial groups revealed meaningful inequities in physics student learning that the simple aggregated model failed to identify.

Our current investigation builds on these findings by examining student learning across social identity groups in science courses that use active engagement instruction. To our knowledge, this investigation's disaggregation across gender, race, ethnicity, first-generation college status, and science discipline provides the most fine-grained, precise quantitative intersectional analysis of STEM student outcomes to date.

**Conceptual framework**

We examined our research questions through a quantitative critical (QuantCrit) framework (Wells & Stage, 2015). Critical theory is commonly associated with qualitative research methodologies (Stage, 2007; Cokley & Awad, 2013). Over the past 16 years, however, QuantCrit has emerged as a way to enact critical research using quantitative methods (Wofford & Winkler, 2022; Garcia *et al.*, 2018; Pearson *et al.*, 2022). Researchers have proposed several tenets to guide QuantCrit research (Gillborn *et al.*, 2018; Castillo & Gillborn, 2022; Covarrubias *et al.*, 2018). In this investigation we forefront 3 tenets: 1) The centrality of oppression, 2) data cannot 'speak for itself' and 3) taking an intersectional perspective.

*The centrality of oppression –*

We take as a fact that structural racism and sexism plague the U.S. economic, political, and educational systems (McGee, 2020. To take a race evasive or neutral stance is to tacitly support the oppressive power structures that create inequity in our society. In this work, we view STEM through the lens of broadening participation theory, which articulates the multitude of mechanisms by which STEM fields fail to recruit, retain, and equally reward underrepresented groups in STEM education throughout the womb-to-workforce continuum (Powell *et al.*, 2018). For example, research has shown Black and Hispanic students have less access to high quality science and mathematics instruction in K-12 schools (Raudenbush, Fotiu & Cheong, 1998; Flores, 2007), and accelerated placement or international baccalaureate courses that can enhance their likelihood of college entry and successful matriculation (Price, 2021). While gender inequity in pre-college achievement has a limited role in STEM trajectories, recent research suggest K-12 experiences lead young women to see their future selves pursuing a career in fields outside of STEM, which ultimately becomes a source of STEM degree and workforce inequity (Weeden, Gelbgiser, & Morgan, 2020).

Higher education courses exacerbate these disparities in STEM diversity. Recent research shows introductory courses disproportionately drive minoritized students out of STEM while leaving C-average White males with a 48% greater likelihood of obtaining a STEM degree than other students (Hatfield, Brown, Topaz, 2022). While informative, these studies either do not focus on intersectional oppression or are limited to race-gender intersectional analyses. Studies with an explicit intersectionality lens tend to examine STEM inequities outside of college degree programs (Vargas-Solar, 2022; Kozlowski, Larivière, Sugimoto & Monroe-White, 2022). One of the most relevant of these analyses documents the privilege of white, heterosexual, able-bodied men in STEM professions (Cech, 2022), which adds additional justification for our adoption of the centrality of oppression framework while using this work as a point of departure to present a more elaborate analysis within the higher education context. Our analysis will include gender, race, ethnicity, and first-generation status within physics, biology, and chemistry degree programs.

*Data cannot 'speak for itself' –*

The practice of reporting data and letting it 'speak for itself' encourages readers to assume that the findings are objective and that they should interpret them from the dominant (e.g., deficit) perspective, which reinforces existing oppressive power structures (Dixon-Roman, 2017). In believing quantitative methods is the appropriate method for centering the collective voice of communities, we take several steps in this investigation to ensure that our findings are not used to undermine the collective voice of marginalized communities with deficit-based interpretations and data inferences. For example, we conceptualize the relative differences in STEM outcomes that stand in the way of achieving broadened participation as society's educational debts (Ladson-Billings, 2006; 2007) and STEM education debts in particular (Jabbari & Johnson, 2022; Nissen *et al.*, 2021). Using an educational debts framing counters implicit assumptions that blame students for these inequities. Contemporary achievement inequalities extend from a history of intentional, systematic, and structural inequities in which dominant groups' decisions and efforts created greater advancement for them than for students whom they minoritized because of their race, gender, family history of no college attendance, or lower household incomes. We explicitly name the sources of society's educational debts measured in this investigation as racism, sexism, and classism.

*Taking an intersectional perspective –*

We rely on intersectionality to account for the multiple social locations that bear on the educational debts that STEM students accumulate (Cech, 2022; Espinosa, 2011). An outgrowth of Black feminist theory, intersectionality posits marginalized women's experiences can be fully understood only when viewed through the lens of multiple systems of oppression, among them race, ethnicity, gender, sexual orientation, and social class (Collins & Bilge, 2020; Crenshaw, 1990). The theory affirms the examination of the intersectional oppression of men of color too (Carbado, 1999), provided such recognition does not elevate their plight over that of women in abolitionist discourse and work (Crenshaw, 2017). Intersectional theory allows us to understand and empirically measure variation in STEM educational debt across combinations of students' social locations and subsequently sharpen the focus of potential remedies for broadening the participation of underrepresented STEM students.

Intersectionality's valuing of individual's unique overlapping social identities is in tension with statistical models' creating groups for analysis. Researchers have proposed several methods for creating quantitative models that best reflect intersectional outcomes. In this analysis, we use an *intracategorical complexity* (McCall, 2005) approach in which we included a saturation of interaction terms between social identity group variables in our models. For example, in addition to having primary terms for first-generation, Black, and women we also include two-way interactions between each as well as a three-way interaction term between them. The inclusion of interaction terms allows the models to create group specific predictions with more accurate point estimates and uncertainty than primary terms alone.

**Positionality Statements**

Including a positionality statement is common for work that takes a critical perspective. Positionality statements acknowledge a researcher's social identity's role in shaping their investigation. Below are the authors' positionality statements.

Ben Van Dusen: I identify as a continuing-generation White cisgender man. I earned a bachelor's degree in physics, have engaged in physic research, was a high school physics teacher, and now have a PhD in education and prepare future science teachers. I was raised in low-income households but now earn an upper-middle-class income. People with similar privileges have created and maintained our society's unjust power structures. As a person with privilege, I believe it is my obligation to use that privilege to dismantle oppressive systems. I also recognize that my privilege, however, limits my perspective on the lived experiences of marginalized individuals.

Jayson Nissen: Identifying as a first-generation, White, cisgendered, man provides me with opportunities denied to others in American society. I have a PhD in physics, teach physics, and worked as a chemist, which focuses my work on the sciences. My experience growing up poor and serving in the all-male submarine service motivated me to reflect on and work to dismantle oppressive power structures in science.

Odis Johnson: I identify as a first-generation college-educated, cis-gender, gay, Black male product of public schools and a working-class neighborhood. My epistemic standpoint in research is as a strategic positivist, and I strive to provide rigorous scientific evidence to uncover the fundamental mechanisms of inequity and inform social transformation. My economic standing, gender, and living in colonized lands require continuous reflection about my privilege.

## Methodology and Methods

**Data Collection**

We accessed the data from the LASSO (Van Dusen, 2018; Nissen *et al.*, 2021) platform database. LASSO is an online assessment platform that administers, scores, and analyses for low-stakes, research-based assessments across the STEM disciplines (Nissen *et al.*, 2018; Van Dusen *et al.*, 2021). The LASSO database provided scores from 18,791 students in 305 courses at 45 institutions (Table 2) on the Introductory Molecular and Cell Biology Assessment (Shi *et al.*, 2010), Chemical Concepts Inventory (Mulford & Robinson, 2002), or the Force Concept Inventory (Hestenes *et al.*, 1992). The student data included their assessment, pretest and posttest scores, social identifiers (gender, race, ethnicity, and first-generation college status), and information about the course structure. None of the instruments showed ceiling or floor effects (Appendix Figure 4).

While the data in this study does not fully represent all institutional contexts, the LASSO database is more representative than the literature (Nissen *et al.*, 2021). Appendix Table 2 details the institution types in the dataset as determined by the Carnegie classification of institutions of higher education (CCIHE) public 2021 database. The appendix includes more information about LASSO and the reliability of its data.

Appendix Table 3 includes the sample size, mean, and standard deviations for the pretest and posttest scores across race, gender, and first-generation status. Appendix Figure 4 plots the scores for the students across the disciplines and social identity groups in the analysis. The figure shows that while there are meaningful differences in average scores between social identity groups there is also significant variation in scores within groups. Despite the impact of racist, sexist, and classist power structures, there are examples of resilient students from each marginalized social identity group who are excelling in their discipline. Appendix Table 1 details the proportion of the pretest sample and the 18-24 year old U.S. population that are men or women for Asian, Black, Hispanic, and White students.

Instructors reported using collaborative learning in 300 of the 305 courses. Appendix Figure 1 shows a breakdown of the specific activities and regularity with which instructors report engaging in them. Physics courses reported the highest use of small group work and lowest rates of lecture. Learning assistants, undergraduates hired to assist in implementing student-centered and small group activities (Otero *et al.*, 2006), supported instruction in 197 of the 305 courses (detailed in Appendix Figure 2).

Our dataset had missing data for four reasons: 1) a question being implemented on LASSO part way through data collection, 2) students opting to not complete either a pretest or posttest, 3) students opting to not include their responses in LASSO's anonymized research database, and 4) students opting to not answer a question on the assessment. The variable with the most missing data was in student first/continuing-generation status (Appendix Table 4) and was primarily due to the question being implemented part way through the data collection process. The variable with the second most missing data was the posttest scores and was primarily due to students not taking the posttest. The rate of missingness of pretest and posttest data is within the normal range for publications in physics education research (Nissen *et al.*, 2019).

**Data Analysis**

We used RStudio (v1.4.1717) (Racine, 2012) to perform our statistical analyses. To clean the data, we removed the pretest or posttest score if the student took less than 5 minutes on the assessment. We removed any courses with less than either nine pretests or nine posttests. To address the missing data and minimize the bias it might cause (Allison, 2001; Buhi *et al.*, 2008; Manly & Wells, 2015; Van Buuren & Groothuis-Oudshoorn, 2011; Woods *et al.*, 2023), we followed the methods that Nissen *et al*. (2019) proposed in their analysis of how best to address missing data from the LASSO database. Specifically, we used the mice package (Woyston and White, 2011) to run a four-level (test, student, course, institution) hierarchical multiple imputation that created ten multiply imputed datasets. We included a 4th level (institutions) to our imputation model because it was a strong predictor of student first/continuing-generation status, which was the variable with the most missing data (Appendix Table 4).

We developed a three-level Bayesian regression model for each instrument. These models nested tests (level 1) in students (level 2) in courses (level 3). We use an *intercategorical* approach to intersectional modeling (McCall, 2005) in which we include interaction terms between all of our gender, race, and first-generation status terms. To ensure model parsimony, we identified our final model as the model with the lowest Akaike information criterion corrected (AICc) (Johnson *et al.*, 2004) calculated by the dredge function in the MuMin package (Bartoń, 2022). We used AICc values (Cavanaugh, 1997) because other common techniques for model selection such as coefficient p-values, additional variance explained, and more restrictive information criterion (e.g., Bayesian information criterion) can eliminate social identifier variables or their interactions even when those variables or interactions represent large differences between groups (Van Dusen & Nissen, 2022). We used the rstan package (Team, 2016) to create our final Bayesian models.

Level-1 equations (assessment-level)
$$Score_{ijk} = \pi_{0jk} + \pi_{1jk} posttest_{ijk} + e_{ijk}$$

Level-2 equations (Student-level)

$$\begin{aligned}\pi_{0jk} = \ &\beta_{00k} + \beta_{01k}\text{Black}_{jk} + \beta_{02k}\text{Hispanic}_{jk} + \beta_{03k}\text{White}_{jk} + \beta_{04k}\text{Hispanic}_{jk} * \text{White}_{jk} \\ &+ Woman_{jk} \\ &* (\beta_{05k} + \beta_{06k}\text{Black}_{jk} + \beta_{07k}\text{Hispanic}_{jk} + \beta_{08k}\text{White}_{jk} + \beta_{09k}\text{Hispanic}_{jk} \\ &* \text{White}_{jk}) + FG_{jk} \\ &* (\beta_{0(10)k} + \beta_{0(11)k}\text{Black}_{jk} + \beta_{0(12)k}\text{Hispanic}_{jk} + \beta_{0(13)k}\text{White}_{jk} \\ &+ \beta_{0(14)k}\text{Hispanic}_{jk} * \text{White}_{jk}) + G_{jk} * Woman_{jk} \\ &* (\beta_{0(15)k} + \beta_{0(16)k}\text{Black}_{jk} + \beta_{0(17)k}\text{Hispanic}_{jk} + \beta_{0(18)k}\text{White}_{jk} \\ &+ \beta_{0(19)k}\text{Hispanic}_{jk} * \text{White}_{jk}) + \beta_{0(20)k}retake_{jk} + \beta_{0(21)k}gender\_other_{jk} \\ &+ \beta_{0(20)k}retake_{jk} + \beta_{0(21)k}gender\_other_{jk} + \beta_{0(22)k}race\_other_{jk} + r_{0jk}\end{aligned}$$

$$\begin{aligned}\pi_{1jk} = \ &\beta_{10k} + \beta_{11k}\text{Black}_{jk} + \beta_{12k}\text{Hispanic}_{jk} + \beta_{13k}\text{White}_{jk} + \beta_{14k}\text{Hispanic}_{jk} * \text{White}_{jk} \\ &+ \text{Woman}_{jk} \\ &* (\beta_{15k} + \beta_{16k}\text{Black}_{jk} + \beta_{17k}\text{Hispanic}_{jk} + \beta_{18k}\text{White}_{jk} + \beta_{19k}\text{Hispanic}_{jk} \\ &* \text{White}_{jk}) + FG_{jk} \\ &* (\beta_{1(10)k} + \beta_{1(11)k}\text{Black}_{jk} + \beta_{1(12)k}\text{Hispanic}_{jk} + \beta_{1(13)k}\text{White}_{jk} \\ &+ \beta_{1(14)k}\text{Hispanic}_{jk} * \text{White}_{jk}) + FG_{jk} * \text{Woman}_{jk} \\ &* (\beta_{1(15)k} + \beta_{1(16)k}\text{Black}_{jk} + \beta_{1(17)k}\text{Hispanic}_{jk} + \beta_{1(18)k}\text{White}_{jk} \\ &+ \beta_{1(19)k}\text{Hispanic}_{jk} * \text{White}_{jk}) + \beta_{1(20)k}retake_{jk} + \beta_{1(21)k}gender\_other_{jk} \\ &+ \beta_{1(22)k}race\_other_{jk} + r_{0jk}\end{aligned}$$

Level-3 equations (Course-level)

$$\begin{aligned}\beta_{00k} &= \gamma_{00k} + u_{00k} \\ \beta_{0(1-22)k} &= \beta_{0(1-22)k} \\ \beta_{1(0-22)k} &= \beta_{1(0-22)k}\end{aligned}$$

Woltamn (2012) provides a detailed description of HLM equations, which we will cover briefly here. The subscripts, for example $Score_{ijk}$, refer to the *i*th assessment in the *j*th student in the kth course. In the level-1 equation, the $\pi_{0jk}$ term represents the score before instruction. The $\pi_{1jk}$ term represents the shift in scores from before to after instruction. The $e_{ijk}$ term represents the assessment-level error for a specific score, is the difference between the predicted and actual values, and is analogous to the $\epsilon$ term in standard linear regressions. In the level-2 equation, the $\beta_{00k}$ term represents the intercept for scores before instruction. The $\beta_{10k}$ represents the intercept for the shift in scores from before to after instruction. The $\beta_{0(1-22)k,1(1-22)k}$ terms are the coefficients for the respective variable in the model. The $r_{0jk}$ term represents the student-level error for each student and allows the intercept to vary across each student. In the level-3 equation, the $\gamma_{00k}$ term is the intercept for the *k*th course. The $\gamma_{0(1-22)k,1(0-22)k}$ terms represent the slopes (e.g., the regression coefficient) for each variable for the *k*th course. The $u_{00k}$ term represents the course-level error and allows the intercept to vary across each course. The model is a fixed slope model since the slopes $\pi_{1jk}$ and $\beta_{0(1-12)k,1(1-12)k}$ equations do not include *r* or *u* variables.

We used visual inspections instead of sensitivity analysis to check the model assumptions because of its computational difficulty (Gelman *et al.*, 1996). Visual inspection showed convergence for all variables.

A major benefit of Bayesian analysis is the ability to use priors to inform a model. Priors are a set of beliefs about the value for coefficients that are included before analyzing the relevant data that can increase the model's accuracy and precision. Priors are usually generated from other related datasets and studies. To set the priors for our physics model, we used the posteriors from a model of student outcomes developed using LASSO data from a similar instrument for introductory physics courses, the Force and Motion Conceptual Evaluation (Thornton & Sokoloff, 1998). The posteriors from the physics model set the priors for the biology and chemistry models. We used the physics model to set the priors for the other models because it had the largest number of courses and the greatest diversity in institution types. Appendix Figure 3 shows the model development workflow with and without priors. The sensitivity analysis showed that including priors had small effects on the estimates and excluding them would not change our conclusions (Appendix Table 5).

When interpreting the uncertainty in our model predictions, we followed the advice of the American Statistical Association (Wasserstein *et al.*, 2019; Wasserstein & Lazar, 2016) in not using p-value cut-off scores. Specifically, we used error bars and larger data trends to evaluate our confidence levels in educational debts. We operationalized society's educational debts as the difference between a marginalized group's predicted performance and that of continuing-generation White men. We contextualized the magnitude of the educational debts by comparing them to the mean gain from pretest to posttest on the instrument to provide a fraction of the learning over a semester (Lortie-Forgues *et al.*, 2021). Our descriptive statistics (Appendix Table 3) includes the mean and standard deviation of the student outcomes.

### Findings

The Findings section focuses on the predicted outcomes for each group (Table 3; Figure 1). We do not directly examine the model's coefficients, which we include in the appendix (Appendix Table 6), because it requires combining up to 32 coefficients to predict a group's score (e.g., First-generation White Hispanic women's posttest scores). To examine society's educational debts before and after instruction (research questions 1 and 2) we first look at each discipline separately. We then examine the trends across the disciplines (research question 3). Except where noted otherwise, the percentages in the findings represent the educational debts in terms of raw differences in assessment scores.Come

### Physics

The educational debts owed to physics students before instruction, shown in Figure 2a, ranged between 1.7% and 20.4%, with a mean of 12.3%. Seven of the nineteen educational debts increased from pre- to post-instruction, with a mean change of -0.6%. The largest decreases in society's education debts during the course were for those debts owed to continuing-generation (-4.6%) and first-generation (-4.1%) White women. The largest increases in society's educational debts during the course were owed to first-generation (5.2%) and continuing-generation (2.3%) Black men. Society owed the largest educational debts after instruction to groups with multiple marginalized identities: first-generation (20.5%) and continuing-generation (20.4%) Hispanic women, first-generation (15.4%) and continuing-generation (17.9%) Black women, and first-generation Black men (17.6%). After instruction, society owed the smallest educational debts to continuing-generation Asian men (2.8%) and first-generation White men (1.4%). These results show that introductory physics courses primarily maintained pre-existing educational debts while increasing the debts owed to Black men and decreasing the debts owed to White women.

### Biology

The educational debts owed to biology students before instruction, shown in Figure 2b, ranged between -0.3% and 7.8%, with a mean of 5.3%. Sixteen of the nineteen educational debts increased from pre- to post-instruction, with a mean increase of 2.5%. The three that decreased were first-generation Hispanic (-1.3%), White (-0.6%), and White Hispanic (-1.4%) men. The largest increases in society's educational debts during the course were owed to first-generation Asian (6.1%) and Black (5.8%) women, continuing-generation Hispanic women (5.3%), and first-generation Black men (5.0%). Society owed the largest educational debts after instruction to first-generation Black men (12.8%) and women (12.6%) and continuing-generation Hispanic women (12.6%). The educational debts owed to first-generation Black men were more than twice as large as any other group of men. After instruction, society owed the smallest educational debts to continuing-generation Asian men (1.6%) and first-generation White men (1.9%). These results show that introductory biology courses meaningfully added to these educational debts, particularly those owed to women and first-generation Black men.

**Chemistry**

The educational debts owed to chemistry students before instruction, shown in Figure 2c, ranged between 0.8% and 15.4%, with a mean of 9.0. None of the nineteen educational debts increased from pre- to post-instruction, with a mean decrease of -4.1%. The largest decreases in society's educational debts during the course were for those owed to continuing-generation (-7.3%) and first-generation (-7.8%) Hispanic women. The smallest changes in society's education debts during the course were for those owed to Black (-0.1%) and White (-0.1%) first-generation men. Society owed the largest educational debts after instruction to first-generation Black men (11.2%), continuing-generation Black women (10%), and continuing-generation White Hispanic women (10.4%). After instruction, society owed the smallest educational debts to continuing-generation (-3.6%) and first-generation (0.1%) Asian men and first-generation White men (0.8%). These results show that introductory chemistry courses meaningfully decreased these educational debts, particularly those owed to Hispanic women.

While these findings indicate collaborative instruction can repay educational debts, the model precision was worse for the chemistry analysis due to a majority (88%) of the first/continuing-generation status data being missing (details are in the appendix). We must temper our claims about the repayment of educational debts to specific groups, particularly around issues of first/continuing-generation status, because of the uncertainty in the predicted outcomes. The trend of Black women, Hispanic women, and first-generation Black men being owed the largest educational debts, however, was consistent across the disciplines.

**Trends**

All models showed strong evidence of meaningful educational debts owed to marginalized students. As discussed in the Methods, we contextualized the magnitude of the educational debts owed to students in each discipline using the mean gain from pretest to posttest as a proxy for the average learning over a course, as shown in Appendix Table 7. The mean gain in physics was 19.6%, biology was 14.8%, and chemistry was 8.6%. The average effect sizes were 0.9 in physics, 0.9 in biology, and 0.4 in chemistry.

Society's average educational debt on the pretest was 12.3% in physics, 5.3% in biology, and 9.0% in chemistry, see Appendix Table 7. When normalized to average amount of learning in a term (as measured by the average gain), there were large differences across the disciplines in society's educational debts on the pretest: 63% of the gain in physics, 36% of the gain in biology, and 104% of the gain in chemistry. Society's average educational debt on the posttest was 11.7% in physics, 7.8% in biology, and 4.9% in chemistry. When normalized to average amount of learning in a term, there were relatively small differences across the disciplines in society's educational debts on the posttest: 60% of the gain in physics, 53% of the gain in biology, and 56% of the gain in chemistry. Physics courses largely maintained society's educational debts, though they decreased for White women and increased for first-generation Black men. Biology courses added to educational debts owed to almost all groups of students, particularly for multiply-marginalized groups. Chemistry courses repaid educational debts to all groups, with the largest repayments for Hispanic women.

Across all three disciplines, the largest educational debts after instruction were consistently owed to Black and Hispanic women and first-generation Black men. The smallest educational debts were consistently owed to Asian men and first-generation White men. The educational debts due to sexism and racism were visible for nearly every group across all three disciplines in Figure 2. The educational debts owed due to classism, however, were small for many groups, but they were consistently larger for Black and Asian men.

**Discussion**

Our research focuses on the extent to which collaborative instruction repays educational debts in conceptual knowledge across the STEM disciplines to broaden participation in the STEM disciplines (Philip & Azevedo, 2017). Scientists often think of science as having a "culture of no culture" (Traweek, 2009; Subramaniam & Wyer, 1998; Shultz *et al*., 2022). The large and persistent educational debts owed to students from multiple marginalized identities in this study, however, show that college science instruction is not only embedded in our society's racist, sexist, and classist power structures but that science education embodies those structures.

Our intersectional perspective provided insight into the ways that racism, sexism, and classism interact to create society's educational debts in the sciences. The results found differences and consistencies across biology, physics, and chemistry. Biology courses increased society's educational debts, physics courses maintained them, and chemistry courses decreased them. Across all three disciplines, society owed large educational debts to women. Classism caused smaller differences in educational debts for women. In all three disciplines, however, society owed first-generation Black and Asian men meaningfully educational debts due to classism. These educational debts indicated that classist systems were particularly detrimental for Black and Asian men in science. Society owed the largest educational debts to Black and Hispanic women and first-generation Black men before and after instruction in all three disciplines.

the physics courses included White men at 2.1 times their proportion of the 18-24 year-old U.S. population (US Census, 2019) whereas the physics courses included Black women and men at 0.4 and 0.5 of their proportions of the 18-24 year-old U.S. population, additional details provided in Appendix Table 1.

The comparisons of conceptual knowledge across races and genders in this study likely underestimates the actual differences due to a 'survivorship bias' (Smith, 2014). The opportunities and pathways to college science courses varies across these groups. These inequities bias our comparisons of who made it into these courses. For example, White men took these chemistry and physics courses at 1.4 and 2.1 times their proportion of the 18-24 year old U.S. population, Appendix Table 1. Whereas they took biology courses at 0.8 times their rate in the population. The students who are owed the largest debts are also the most likely to be denied access to the courses. This study likely underestimates the true educational debts in chemistry and physics more so than in biology. Given that the biology courses were more diverse, our finding that society's educational debts were smaller than in chemistry and physics is particularly noteworthy. Identifying features of our K-12 education system that create more diverse representation and smaller educational debts in biology can inform how to replicate these features in chemistry and physics. Metcalf (2016), however, also points out that aggregating across the many distinct fields in university biology courses can obscure inequities within those fields.

The change in educational debts society owed to marginalized students was mixed over the semester. Despite students in biology courses being owed the smallest educational debts before instruction, the debts were exacerbated during the course. The opposite trend was seen in chemistry courses, which began with the largest societal educational debts (measured as a share of the learning over a semester) but mitigated them during the course. Physics courses maintained society's educational debts. After instruction, society's educational debts owed to marginalized students as a fraction of the learning in a course were relatively similar across the disciplines (53-60%). Our data does not clarify why the biology courses exacerbated societal educational debts while the chemistry courses mitigated them. The results, however, do indicate that instructors spending more time on collaborative activities is not sufficient to repay educational debts.

While Theobald *et al*. (2020) found that using more collaborative learning reduced inequities in student outcomes compared to lecture-based instruction. Our investigation found that collaborative instruction courses can maintain and increase educational debts. Despite most courses in our dataset reportedly using collaborative learning regularly, we do not see large-scale repayment of society's educational debts. The physics courses reported the least use of lecture and most use of small group work Appendix Figure 1, yet they maintained society's educational debts. Abundant research shows that collaborative learning in STEM courses improves student outcomes (National Research Council, 2012; Freeman *et al*., 2014) and will provide a foundation for developing pedagogies and curriculum that repay society's educational debts. Explicitly anti-oppressive pedagogical practices and curriculum can support STEM courses in mitigating society's educational debts (Johnson *et al.*, 2020; Ernest *et al*., 2019; Higgins *et al.*, 2018).

Another factor that our data shows to be insufficient for mitigating society's educational debts is equitable gender representation. Women were overrepresented in biology courses, yet the educational debts owed to them increased through instruction. The fact that women are well represented in chemistry and biology is an important step toward the goal of broadening participation. Broadening participation alone, however, does not ensure equitable outcomes.

To create large-scale changes in STEM instructional practices, colleges and science departments must make meaningful changes to support broadening participation (Hatfield *et al.,* 2022). Broadening participation will require that we dismantle the structures that create inequitable representation of marginalized students before and during college science courses. The National Science Foundation has integrated the goals of broadening participation into its solicitations. Most institutions, however, have no analogous mechanisms when reviewing instructors' job performance (e.g., annual review or promotion and tenure). The failure to emphasize creating equitable learning environments may result from a lack of tools to measure and analyze student outcomes. Some institutions (e.g., the California State University system) have begun creating dashboards that allow instructors and administrators to easily visualize student outcomes along many dimensions, including examining gender and racial inequities. Institutions can also draw on free online resources, such as EQUIP (Reinholz & Shah, 2018; Reinholz *et al*., 2022) and the Learning About STEM Student Outcomes (LASSO) platform (Van Dusen, 2018; Nissen *et al*. 2021) used in this investigation, that can automate data collection and equity analyses. These kinds of tools create equity reports that can fuel departmental conversations and instructional transformation.

Another avenue of transformation for STEM departments consistent with broadening participation theory is through collaborations with minority-serving institutions (e.g., Historically Black Colleges and Universities, Hispanic Serving Institutions, and Tribal Colleges and Universities). Institutions founded specifically to serve marginalized populations have extensive expertise in creating more equitable student outcomes. For example, HBCUs only make up 3% of institutions in the United States, yet produce almost 20% of all African American graduates (United Negro College Fund, 2022). Many institutions would benefit from creating opportunities for their instructors to learn from instructors at minority-serving institutions. When developing these partnerships, engaging with instructors from minority-serving institutions as equal partners is important, as is compensating them for their time and expertise (De Leone *et al*., 2019).

We do not know of any other large-scale publication on higher education STEM student outcomes that uses an intersectional perspective to disaggregate data in as many ways as this investigation (gender, race, ethnicity, and first-generation status). This is partially due to the difficulty of collecting sufficient data to power this sort of analysis. There are several things the field can do to lower barriers to engaging in this work. First, support large-scale data collection platforms (e.g., LASSO). This support comes from instructors and researchers using the platform and funding agencies' financial support. Second, researchers can engage in open science practices (Foster & Deardorff, 2017; Banks *et al*., 2019). This includes open-sourcing datasets and reporting descriptive statistics in ways that allow them to be used in meta-analyses. Third, create collaborations within and across institutions, similar to the Sloan Equity and Inclusion in STEM Introductory Courses project (McKay, 2020), that pool resources, create accountability for action, leverage lived-experiences and expertise, and hold those with power accountable to action. Increasing access to large-scale data will power future intersectional investigations that can identify instructional practices that support more equitable STEM student outcomes and broaden participation in STEM. Administrators and community members can support this research, leading to programs and practices that broaden participation by redressing societal injustices.

University science instruction is critical in addressing inequity in science student outcomes. To broaden participation, universities, departments, and instructors must acknowledge that society owes large educational debts to many students entering their courses and take on the responsibility to repay those debts. Ignoring these educational debts and our obligation to repay them makes us complicit in replicating the unjust system that created them.

## Acknowledgments

This work was made possible by the LASSO platform. We thank Sayali Kukday for her valuable feedback on this manuscript. This work is supported, in part, by funds from the National Science Foundation (NSF # 1928596).

## Declaration of Interest Statement

No potential conflict of interest was reported by the author(s).

15
Society's educational debts in science

17
Society's educational debts in science

## Appendix

The following section provides additional details on the analyses. Appendix Table 6 details the model coefficients and their standard errors for the three models. Appendix Figure 4 provides violin, box, and scatter plots for the pretest and posttest scores for each of the ten social identity groups using the imputed data.

**Learning About STEM Student outcomes (LASSO)**

The LASSO database includes anonymized student- and course-level information from STEM courses at over 120 institutions of higher education (Van Dusen, 2018). Several studies have examined the reliability of collecting data online using low-stakes, research-based assessments in the STEM disciplines (Nissen *et al.*, 2018; Bonham, 2008; Wilcox & Pollock, 2019). Using a randomized control trial design, Nissen *et al*. (2018) found that student scores collected with LASSO were similar to scores collected on paper in class. Nissen and colleagues also found that instructors could achieve similar participation levels if they provided participation credit and in-class and email reminders to complete the assessments.

**Table 1**

*Definition of some statistical modeling and equity-related terms we use in the manuscript.*

| Term | Definition |
|---|---|
| Hierarchical Linear Model | A linear regression model that accounts for the nested nature of data (e.g., students within courses). Also known as a multilevel model, linear mixed model, or mixed-effects model. |
| Bayesian | A statistical system based on probability rather than frequency. Bayesian confidence intervals are a measure of P(hypothesis\|data) rather than the frequentist measure of P(data\|hypothesis). |
| Multiple Imputation | A principled approach to handling missing data that generates multiple complete data sets based on collected data and combines the results to account for the increased variance. |
| Equality | When individuals or groups have the same resources and opportunities and achieve the same outcomes. In this work, we focus on equal outcomes. |
| Equity | When a course allocates resources and opportunities according to each person's circumstances to reach equal outcomes. |
| QuantCrit | A theoretical framework that applies critical race theory to quantitative research. |
| Broadening Participation | Increasing participation from underrepresented groups and diverse institutions in the STEM disciplines. Also, increasing marginalized individuals and community agency by recognizing the many ways STEM is valuable to them in career choices and everyday life. |
| Educational Debt | ``Education debt is the foregone schooling resources that we could have (should have) been investing in (primarily) low-income kids, which deficit leads to a variety of social problems (e.g., crime, low productivity, low wages, low labor force participation) that require on-going public investment.'' This reframes inequities in group outcomes from deficits in student abilities to debts that society owes marginalized students due to racism, sexism, and classism Ladson-Billings (2006; 2007). |
| Intersectionality | Intersectionality asserts that the power relations of race, class, and gender, for example, are not discrete and mutually exclusive entities but rather build on each other and work together (Collins & Bilge, 2020). |
| First-Generation | A student whose parent(s) or guardian(s) do not have a bachelor's degree. |

Society's educational debts in science

**Table 2.**

*The number of students, courses, and institutions in the sample disaggregated by discipline. Courses were also disaggregated to indicate the use of collaborative learning or Learning Assistants. Note: Some students and institutions were represented in multiple disciplines making the total less than the sum of the individual values.*

| Discipline | Students | Courses | | | Institutions |
| --- | --- | --- | --- | --- | --- |
| | | Total | Collaborative Learning | Learning Assistant | |
| Biology | 8,305 | 97 | 95 | 53 | 11 |
| Chemistry | 4,576 | 37 | 34 | 31 | 12 |
| Physics | 5,955 | 171 | 171 | 113 | 30 |
| **Total** | **18,791** | **305** | **300** | **197** | **44** |

**Table 3.**
Predicted pretest and posttest scores from our Bayesian model disaggregated by race, first-generation status, and gender.

| Social Identifiers | | | Physics (FCI) | | | | Biology (IMCA) | | | | Chemistry (CCI) | | | |
|---|---|---|---|---|---|---|---|---|---|---|---|---|---|---|
| | | | Pretest (%) | | Posttest (%) | | Pretest (%) | | Posttest (%) | | Pretest (%) | | Posttest (%) | |
| Race | Class | Gender | Score | SE | Score | SE | Score | SE | Score | SE | Score | SE | Score | SE |
| Asian | First-generation | Women | 37.8 | 2.6 | 58.7 | 3.3 | 46.2 | 1.3 | 57.3 | 1.8 | 45.3 | 1.7 | 53.9 | 2.2 |
| | | Men | 44.4 | 1.3 | 63.2 | 2.1 | 47.3 | 1.1 | 63.0 | 1.9 | 48.8 | 1.4 | 58.2 | 1.8 |
| | Continuing-generation | Women | 37.9 | 1.7 | 59.9 | 2.1 | 46.5 | 0.9 | 59.5 | 2.2 | 44.7 | 1.4 | 54.9 | 3.7 |
| | | Men | 47.7 | 1.5 | 65.8 | 1.7 | 49.6 | 0.9 | 64.9 | 1.3 | 51.6 | 1.4 | 61.9 | 1.7 |
| Black | First-generation | Women | 32.8 | 3.1 | 53.1 | 3.8 | 42.6 | 1.6 | 53.9 | 2.4 | 40.3 | 2.9 | 49.7 | 3.3 |
| | | Men | 37.4 | 2.2 | 51.0 | 2.6 | 41.5 | 1.9 | 53.7 | 2.4 | 41.2 | 2.0 | 47.1 | 3.3 |
| | Continuing-generation | Women | 30.3 | 2.0 | 50.6 | 2.6 | 42.1 | 1.3 | 54.9 | 2.8 | 38.0 | 1.8 | 48.3 | 3.6 |
| | | Men | 41.8 | 1.9 | 58.2 | 2.4 | 45.7 | 1.5 | 60.7 | 2.0 | 45.7 | 1.9 | 55.0 | 2.5 |
| Hispanic | First-generation | Women | 30.6 | 2.9 | 48.1 | 4.3 | 42.0 | 1.7 | 57.2 | 3.4 | 38.8 | 3.3 | 52.6 | 5.2 |
| | | Men | 36.4 | 1.7 | 54.5 | 3.0 | 42.7 | 1.8 | 61.2 | 2.3 | 42.9 | 2.3 | 53.3 | 3.9 |
| | Continuing-generation | Women | 29.4 | 2.3 | 48.2 | 2.5 | 41.9 | 1.7 | 53.8 | 3.4 | 38.5 | 2.1 | 51.6 | 3.0 |
| | | Men | 38.0 | 1.6 | 56.0 | 2.2 | 43.2 | 1.3 | 59.7 | 2.0 | 44.6 | 1.7 | 56.6 | 3.0 |
| White | First-generation | Women | 36.2 | 1.4 | 59.0 | 3.2 | 43.1 | 1.1 | 56.2 | 2.1 | 42.2 | 1.3 | 53.2 | 2.3 |
| | | Men | 48.2 | 1.2 | 67.1 | 2.4 | 46.8 | 1.6 | 64.5 | 1.7 | 51.5 | 2.6 | 57.5 | 2.2 |
| | Continuing-generation | Women | 36.0 | 1.3 | 59.2 | 2.8 | 43.3 | 0.8 | 58.1 | 2.1 | 41.4 | 1.2 | 51.0 | 3.0 |
| | | Men | 49.8 | 1.6 | 68.6 | 2.0 | 49.3 | 1.1 | 66.5 | 1.5 | 52.4 | 1.6 | 58.3 | 2.0 |
| White Hispanic | First-generation | Women | 32.8 | 2.8 | 53.9 | 3.0 | 41.5 | 1.4 | 57.5 | 1.4 | 37.9 | 2.7 | 49.8 | 4.1 |
| | | Men | 40.3 | 2.2 | 61.1 | 2.5 | 42.8 | 1.9 | 61.4 | 2.8 | 47.0 | 3.2 | 56.8 | 3.9 |
| | Continuing-generation | Women | 33.3 | 2.1 | 53.0 | 3.2 | 42.2 | 1.3 | 56.5 | 2.3 | 37.0 | 2.0 | 47.9 | 3.3 |
| | | Men | 41.3 | 2.2 | 59.7 | 1.9 | 44.0 | 1.5 | 60.2 | 2.1 | 47.2 | 1.8 | 56.0 | 3.1 |

**Appendix Table 1**
*Proportions of the pretest sample and 18-24 year old U.S. Population. Relative proportion is the percentage that a group is over- or under-represented in the dataset relative to the U.S. population. We combined the non-White and White Hispanic students into one group to compare to data from the U.S. census (US Census, 2019).*

| Race | Gender | Study Participants on Pretest | | | U.S. Population 18-24 years old (%) | Relative Proportion | | |
|---|---|---|---|---|---|---|---|---|
| | | Physics (%) | Biology (%) | Chemistry (%) | | Physics (%) | Biology (%) | Chemistry (%) |
| Asian | Women | 6 | 15 | 9 | 3 | 217 | 503 | 308 |
| | Men | 10 | 7 | 8 | 3 | 349 | 231 | 255 |
| Black | Women | 3 | 8 | 5 | 8 | 40 | 101 | 67 |
| | Men | 4 | 2 | 3 | 8 | 49 | 26 | 34 |
| Hispanic | Women | 6 | 12 | 7 | 11 | 53 | 108 | 68 |
| | Men | 14 | 5 | 5 | 12 | 119 | 40 | 45 |
| White | Women | 21 | 43 | 47 | 26 | 79 | 166 | 179 |
| | Men | 56 | 19 | 37 | 27 | 208 | 72 | 136 |

**Appendix Table 2.**
*Institution information from our dataset. Note: The subcategories do not add up the total because 2 institutions were not in the Carnegie classification of institutions of higher education (CCIHE) public 2021 database.*

| Total | Type | | Size | | | Highest Degree | | | | Special Designators | |
|---|---|---|---|---|---|---|---|---|---|---|---|
| | Public | Private | Small | Medium | Large | AA | BA | MA | PhD | Hispanic-Serving Institution | Minority-Serving Institution |
| 45 | 30 | 13 | 5 | 13 | 25 | 3 | 5 | 13 | 22 | 8 | 8 |

**Appendix Table 3.**
*Descriptive statistics disaggregated by instrument and social identifiers.*

| | | | Physics (FCI) | | | | | | Biology (IMCA) | | | | | | Chemistry (CCI) | | | | | |
|---|---|---|---|---|---|---|---|---|---|---|---|---|---|---|---|---|---|---|---|---|
| | Social Identifiers | | Pretest (%) | | | Posttest (%) | | | Pretest (%) | | | Posttest (%) | | | Pretest (%) | | | Posttest (%) | | |
| Race | Class | Gender | N | Score | SD | N | Score | SD | N | Score | SD | N | Score | SD | N | Score | SD | N | Score | SD |
| All | All | All | 4726 | 44.1 | 22 | 4005 | 64.8 | 23 | 7162 | 45.3 | 17 | 5685 | 60.3 | 18 | 3637 | 47.2 | 19 | 2233 | 52.6 | 21 |
| Asian | First-generation | Women | 105 | 39.6 | 22.0 | 96 | 58.8 | 26.8 | 485 | 46.5 | 17.6 | 433 | 60.6 | 17.8 | 133 | 43.2 | 20.2 | 115 | 60.2 | 23.7 |
| | | Men | 252 | 45.3 | 22.4 | 229 | 67.6 | 21.4 | 162 | 51.8 | 20.4 | 150 | 62.3 | 16.8 | 157 | 49.9 | 18.4 | 149 | 54.8 | 23.0 |
| | Continuing-generation | Women | 202 | 38.9 | 19.2 | 210 | 62.9 | 21.8 | 596 | 48.4 | 18.2 | 646 | 59.0 | 16.6 | 203 | 49.1 | 19.1 | 220 | 58.9 | 21.4 |
| | | Men | 243 | 51.1 | 25.3 | 265 | 66.3 | 21.1 | 334 | 48.1 | 18.8 | 344 | 70.1 | 16.2 | 121 | 55.2 | 21.1 | 129 | 57.1 | 21.5 |
| Black | First-generation | Women | 71 | 30.5 | 20.2 | 65 | 49.3 | 22.8 | 201 | 40.4 | 16.1 | 218 | 54.0 | 18.9 | 100 | 33.9 | 16.1 | 94 | 45.5 | 21.6 |
| | | Men | 76 | 40.9 | 21.2 | 78 | 53.2 | 22.6 | 42 | 40.9 | 17.4 | 44 | 62.4 | 20.1 | 62 | 41.3 | 16.6 | 51 | 47.8 | 17.7 |
| | Continuing-generation | Women | 80 | 32.3 | 20.0 | 86 | 48.5 | 20.3 | 378 | 43.5 | 16.3 | 361 | 54.5 | 18.9 | 96 | 37.9 | 19.4 | 104 | 40.1 | 20.5 |
| | | Men | 108 | 40.8 | 25.1 | 108 | 55.4 | 20.6 | 109 | 43.6 | 21.1 | 108 | 65.3 | 18.5 | 36 | 42.4 | 13.8 | 47 | 50.8 | 19.2 |
| Hispanic | First-generation | Women | 46 | 33.8 | 21.8 | 47 | 54.6 | 23.6 | 125 | 39.6 | 16.1 | 121 | 60.3 | 20.9 | 39 | 35.4 | 17.2 | 53 | 60.2 | 15.5 |
| | | Men | 102 | 37.9 | 17.0 | 86 | 57.2 | 18.7 | 53 | 42.2 | 17.5 | 47 | 62.4 | 19.1 | 19 | 38.7 | 25.5 | 19 | 58.4 | 19.5 |
| | Continuing-generation | Women | 75 | 27.9 | 20.3 | 73 | 48.7 | 22.9 | 150 | 40.7 | 16.5 | 154 | 57.6 | 20.3 | 59 | 43.9 | 16.6 | 45 | 53.7 | 20.3 |
| | | Men | 207 | 35.8 | 17.5 | 223 | 56.6 | 18.9 | 64 | 38.4 | 19.5 | 71 | 63.6 | 16.6 | 43 | 50.1 | 18.9 | 43 | 51.0 | 21.5 |
| White | First-generation | Women | 290 | 39.6 | 19.5 | 276 | 59.4 | 25.6 | 823 | 43.4 | 17.3 | 786 | 54.4 | 18.2 | 519 | 42.1 | 18.1 | 502 | 52.7 | 18.2 |
| | | Men | 544 | 49.1 | 23.1 | 571 | 66.9 | 19.6 | 259 | 42.0 | 17.9 | 251 | 70.6 | 17.2 | 445 | 53.5 | 15.7 | 438 | 54.9 | 17.7 |
| | Continuing-generation | Women | 682 | 40.1 | 18.8 | 697 | 62.6 | 25.8 | 2268 | 44.2 | 16.4 | 2303 | 55.5 | 18.3 | 1175 | 42.8 | 17.9 | 1191 | 56.5 | 20.5 |
| | | Men | 2110 | 52.1 | 22.3 | 2088 | 67.7 | 19.8 | 1125 | 48.2 | 19.5 | 1135 | 69.6 | 16.1 | 887 | 50.7 | 18.9 | 893 | 52.7 | 20.0 |
| White Hispanic | First-generation | Women | 63 | 31.2 | 14.6 | 60 | 52.8 | 21.4 | 267 | 42.9 | 17.2 | 260 | 57.5 | 16.8 | 76 | 37.9 | 17.7 | 77 | 47.7 | 19.6 |
| | | Men | 169 | 38.0 | 16.6 | 161 | 55.8 | 22.5 | 86 | 44.2 | 19.2 | 87 | 58.7 | 19.4 | 53 | 45.8 | 18.5 | 52 | 53.4 | 18.8 |
| | Continuing-generation | Women | 94 | 36.7 | 20.3 | 98 | 52.0 | 20.5 | 312 | 42.1 | 15.9 | 319 | 60.4 | 15.9 | 98 | 36.1 | 15.6 | 97 | 53.7 | 20.6 |
| | | Men | 197 | 44.0 | 22.6 | 207 | 63.3 | 20.6 | 142 | 43.3 | 18.6 | 140 | 68.7 | 15.3 | 81 | 52.1 | 16.0 | 82 | 50.6 | 19.0 |

**Appendix Table 4.**
Percentage of the data that was missing for each variable disaggregated by instrument.

| Assessment | N | Pretest (%) | Posttest (%) | Gender (%) | Race (%) | FG status (%) |
|---|---|---|---|---|---|---|
| FCI | 17016 | 17.5 | 31.8 | 0.3 | 0.3 | 62.2 |
| IMCA | 8503 | 11.8 | 30.3 | 0.3 | 0.3 | 60.7 |
| CCI | 4714 | 18.1 | 49.2 | 0.2 | 0.2 | 87.8 |

**Appendix Table 5.**
Shifts in our model's predicted society's educational debts when adding informed priors. The shift in mean scores ranged from -2.6 to 2.4. The shift in mean standard errors ranged from 0.5 to -3.3.

| Social Identifiers | | | Physics (FCI) | | | | Biology (IMCA) | | | | Chemistry (CCI) | | | |
|---|---|---|---|---|---|---|---|---|---|---|---|---|---|---|
| | | | Pretest (%) | | Posttest (%) | | Pretest (%) | | Posttest (%) | | Pretest (%) | | Posttest (%) | |
| Race | Class | Gender | Score | SE | Score | SE | Score | SE | Score | SE | Score | SE | Score | SE |
| Asian | First-generation | Women | 1.1 | -0.6 | 0.1 | 0.0 | 0.2 | -0.2 | -0.2 | 0.0 | -0.2 | -0.9 | 0.4 | -0.7 |
| | | Men | -1.1 | -0.2 | -0.5 | -0.1 | 2.2 | -1.0 | -0.3 | -0.5 | 1.2 | -0.9 | -0.7 | -1.6 |
| | Continuing-generation | Women | -0.7 | -0.9 | -0.4 | -0.5 | 1.1 | -0.5 | -0.3 | -0.2 | 0.7 | -0.9 | -0.8 | -1.0 |
| | | Men | 0.3 | -1.7 | 1.0 | -0.4 | -0.4 | -1.4 | -1.4 | -0.6 | 0.6 | -2.2 | -2.5 | -0.7 |
| Black | First-generation | Women | 0.2 | -0.3 | 0.2 | -0.1 | 0.5 | -0.4 | -0.1 | -0.2 | 0.3 | -0.9 | 0.6 | -0.2 |
| | | Men | 0.0 | -0.4 | -0.3 | -0.2 | 0.2 | -1.6 | 0.1 | -0.7 | -1.4 | -1.2 | 1.5 | -1.2 |
| | Continuing-generation | Women | 0.1 | -0.4 | 0.1 | -0.2 | 0.7 | -0.2 | 0.1 | -0.3 | 0.0 | -1.0 | 1.1 | -0.8 |
| | | Men | 0.2 | -0.7 | 0.2 | -0.6 | 1.5 | -1.1 | 0.5 | -0.4 | 0.0 | -3.3 | -0.3 | -1.9 |
| Hispanic | First-generation | Women | 0.5 | -0.4 | -0.4 | -0.1 | 0.2 | -0.2 | 0.2 | -0.2 | -1.8 | -1.2 | 1.8 | -0.6 |
| | | Men | -0.3 | -0.4 | 0.5 | -0.3 | 0.8 | -0.5 | 0.4 | -0.1 | 0.7 | -2.9 | -0.5 | -0.4 |
| | Continuing-generation | Women | -2.6 | -0.6 | 0.2 | -0.2 | 1.4 | -0.6 | -0.3 | -0.6 | 1.4 | -0.7 | 1.7 | -0.8 |
| | | Men | 1.2 | -0.6 | 0.1 | -0.2 | 0.8 | -1.4 | 1.9 | -1.1 | 2.3 | -3.1 | 1.5 | -2.2 |
| White | First-generation | Women | -0.2 | 0.0 | 0.0 | -0.1 | 0.6 | -0.1 | 0.0 | -0.1 | -0.2 | -0.2 | 0.9 | -0.2 |
| | | Men | 0.2 | -0.2 | 0.1 | 0.0 | -0.4 | -1.1 | 0.2 | -0.1 | -0.3 | -0.9 | 1.2 | -0.3 |
| | Continuing-generation | Women | -0.1 | -0.3 | 0.1 | -0.1 | 0.5 | -0.1 | 0.0 | -0.1 | 0.0 | -0.4 | 0.9 | -0.2 |
| | | Men | 0.0 | 0.0 | 0.0 | 0.0 | 0.0 | 0.0 | 0.0 | 0.0 | 0.0 | 0.0 | 0.0 | 0.0 |
| White Hispanic | First-generation | Women | -0.1 | -0.3 | 0.2 | -0.1 | 0.5 | -0.3 | 0.0 | 0.0 | 0.4 | -0.2 | 0.6 | -0.4 |
| | | Men | -0.1 | -0.1 | -0.1 | -0.1 | 0.9 | -0.5 | -0.1 | -0.2 | 0.1 | -1.0 | 1.1 | -1.2 |
| | Continuing-generation | Women | 1.0 | -0.3 | 0.0 | -0.1 | 0.2 | -0.2 | 0.1 | -0.1 | -1.2 | -0.4 | 0.3 | -0.3 |
| | | Men | -0.4 | -0.6 | -0.1 | -0.1 | 1.3 | -0.5 | -0.5 | -0.7 | 0.7 | -1.1 | 2.4 | -1.1 |

**Appendix Table 6.**
Model coefficients and estimated error.

| Coefficient | Physics | | Biology | | Chemistry | |
|---|---|---|---|---|---|---|
| | Estimate | Est.Err. | Estimate | Est.Err. | Estimate | Est.Err. |
| Intercept | 47.68 | 1.62 | 49.57 | 1.07 | 51.56 | 1.47 |
| gender_other | 2.55 | 1.52 | 1.22 | 1.25 | 2.41 | 1.41 |
| race_other | -4.26 | 1.45 | -1.56 | 1.33 | -1.06 | 1.41 |
| retake | 0.81 | 0.99 | -0.23 | 0.70 | -1.19 | 0.86 |
| test | 18.11 | 1.44 | 15.30 | 1.48 | 10.33 | 1.18 |
| FG | -3.31 | 1.33 | -2.31 | 1.07 | -2.72 | 1.07 |
| women | -9.75 | 2.02 | -3.05 | 0.91 | -6.89 | 1.55 |
| black | -5.91 | 1.69 | -3.87 | 1.14 | -5.89 | 1.37 |
| hispanic | -9.64 | 1.95 | -6.33 | 1.34 | -6.92 | 1.44 |
| white | 2.14 | 1.16 | -0.28 | 0.93 | 0.84 | 0.99 |
| test*FG | 0.70 | 1.95 | 0.47 | 1.68 | -0.98 | 1.60 |
| test*women | 3.88 | 2.88 | -2.35 | 2.05 | -0.12 | 3.25 |
| FG*women | 3.19 | 2.39 | 1.99 | 1.42 | 3.35 | 1.64 |
| hispanic*white | 1.10 | 2.10 | 1.00 | 1.35 | 1.69 | 1.62 |
| test*black | -1.70 | 2.07 | -0.31 | 1.51 | -0.99 | 1.83 |
| test*hispanic | -0.17 | 2.00 | 1.13 | 2.06 | 1.63 | 2.67 |
| test*white | 0.61 | 2.05 | 1.87 | 1.32 | -4.44 | 2.36 |
| FG*black | -1.06 | 2.47 | -1.89 | 1.89 | -1.76 | 2.02 |
| FG*hispanic | 1.62 | 2.59 | 1.75 | 1.62 | 1.01 | 2.25 |
| FG*white | 1.65 | 1.96 | -0.18 | 2.15 | 1.84 | 1.63 |
| women*black | -1.75 | 2.81 | -0.59 | 1.43 | -0.81 | 1.88 |
| women*hispanic | 1.15 | 2.92 | 1.74 | 1.81 | 0.71 | 2.09 |
| women*white | -4.12 | 2.06 | -2.98 | 1.36 | -4.07 | 1.43 |
| test*FG*women | -1.84 | 3.43 | -2.34 | 2.47 | -0.60 | 4.36 |
| test*hispanic*white | -0.15 | 2.73 | -2.09 | 2.32 | 1.25 | 2.37 |
| FG*hispanic*white | -0.97 | 3.38 | -0.39 | 2.32 | -0.34 | 2.72 |
| test*FG*black | -3.54 | 3.55 | -3.29 | 2.64 | -2.40 | 3.14 |
| test*FG*hispanic | -0.47 | 4.52 | 1.62 | 2.99 | -0.57 | 4.08 |
| test*FG*white | -0.46 | 2.12 | 0.10 | 1.50 | 1.09 | 2.13 |
| women*hispanic*white | 4.78 | 3.63 | 2.56 | 2.13 | 0.09 | 2.41 |
| test*women*black | 0.08 | 3.87 | 0.20 | 3.52 | 1.15 | 3.67 |
| test*women*hispanic | -3.10 | 3.42 | -2.17 | 4.16 | 1.31 | 4.57 |
| test*women*white | 0.68 | 4.16 | 0.04 | 3.03 | 3.80 | 4.02 |
| FG*women*black | 3.73 | 5.06 | 2.73 | 2.99 | 3.42 | 3.58 |
| FG*women*hispanic | -0.37 | 4.20 | -1.37 | 3.05 | -1.27 | 3.41 |
| FG*women*white | -1.30 | 2.83 | 0.33 | 1.88 | -1.75 | 2.32 |
| test*FG*hispanic*white | 2.59 | 4.95 | 0.22 | 3.36 | 1.47 | 3.63 |
| test*women*hispanic*white | -0.19 | 4.09 | 2.51 | 4.59 | -2.86 | 3.73 |
| FG*women*hispanic*white | -1.03 | 6.14 | -0.53 | 3.58 | 0.78 | 4.86 |
| test*FG*women*black | 4.60 | 6.33 | 3.64 | 5.32 | 3.02 | 5.03 |
| test*FG*women*hispanic | 0.38 | 5.79 | 3.58 | 4.95 | 2.74 | 5.15 |
| test*FG*women*white | 1.12 | 4.54 | 0.00 | 2.56 | 1.93 | 3.49 |
| test*FG*women*hispanic*white | -0.60 | 8.41 | -1.88 | 6.21 | -4.12 | 7.28 |

**Appendix Table 7.**
Mean gain, standard deviation, effect size (*d*), educational debt before instruction, and educational debt after instruction by discipline.

| Discipline | Mean gain (%) | S.D. (%) | *d* | Debt$_{pre}$ (%) | Debt$_{post}$ (%) |
|---|---|---|---|---|---|
| Physics | 19.6 | 21.7 | 0.9 | 12.3 | 11.7 |
| Biology | 14.8 | 17.2 | 0.9 | 5.3 | 7.8 |
| Chemistry | 8.6 | 19.4 | 0.4 | 9.0 | 4.9 |

*Figure 1. Predicted pretest and posttest scores disaggregated by social identifiers on the (a) Force Concept Inventory (FCI), (b) Introductory Molecular and Cell Biology Assessment (IMCA), and (c) Chemical Concepts Inventory (CCI). Error bars represent +/- 1 standard error.*

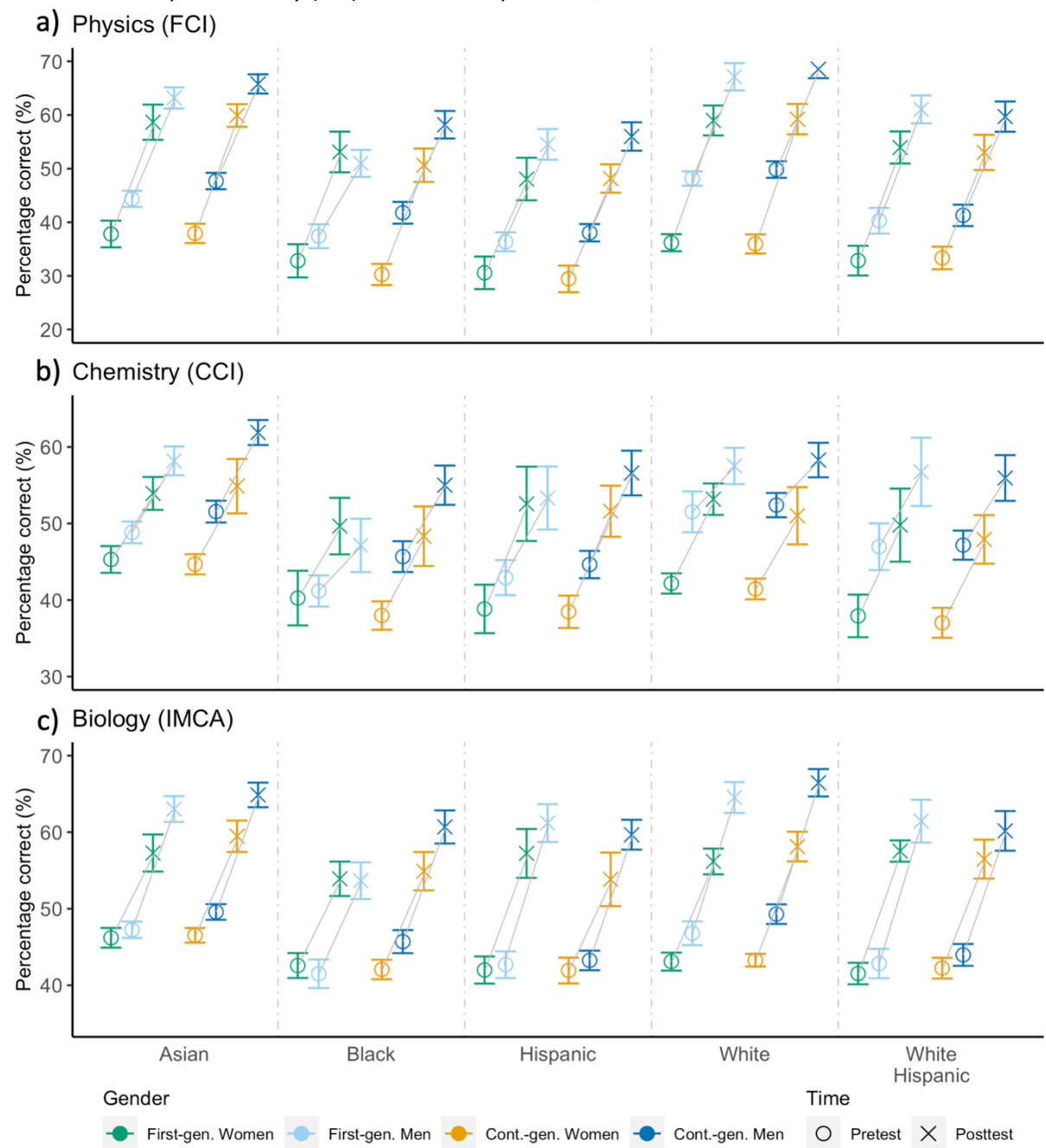

*Figure 2. The educational debts society owed introductory college students as measured by the difference in predicted assessment performance between marginalized groups and continuing-generation White men on the pretest and posttest for (a) physics, (b) biology, and (c) chemistry. Positive values indicate lower predicted scores for marginalized groups. Lines connect the pretest to the posttest for ease of comparison.*

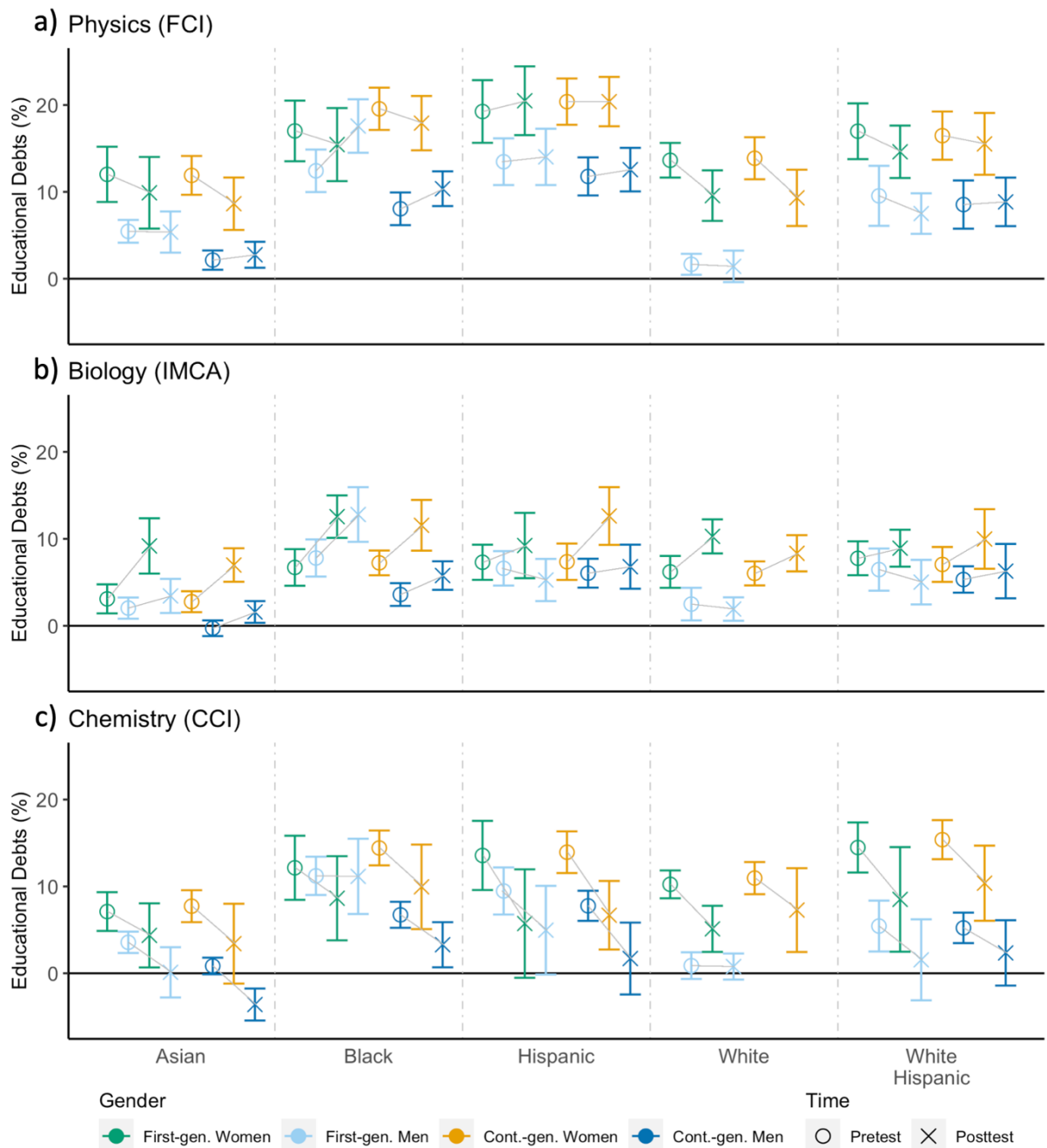

*Appendix Figure 1. A breakdown of the specific classroom activities and regularity with which instructors report engaging in them.*

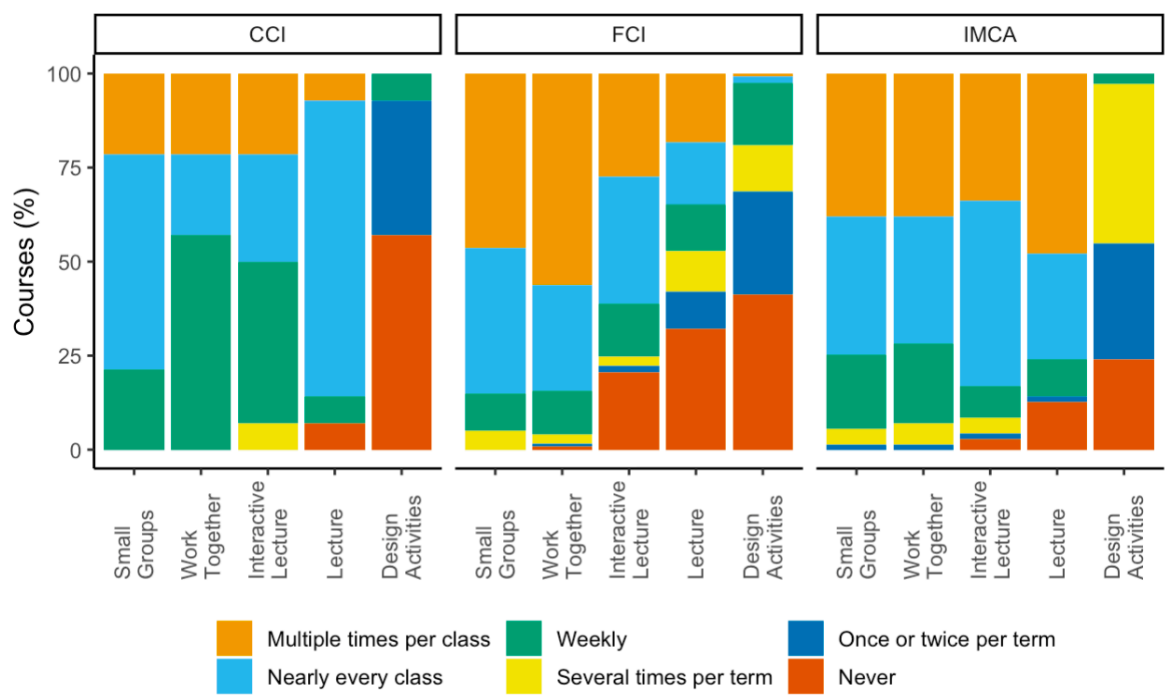

Society's educational debts in science

*Appendix Figure 2. A breakdown of the primary and secondary activities the learning assistants supported.*

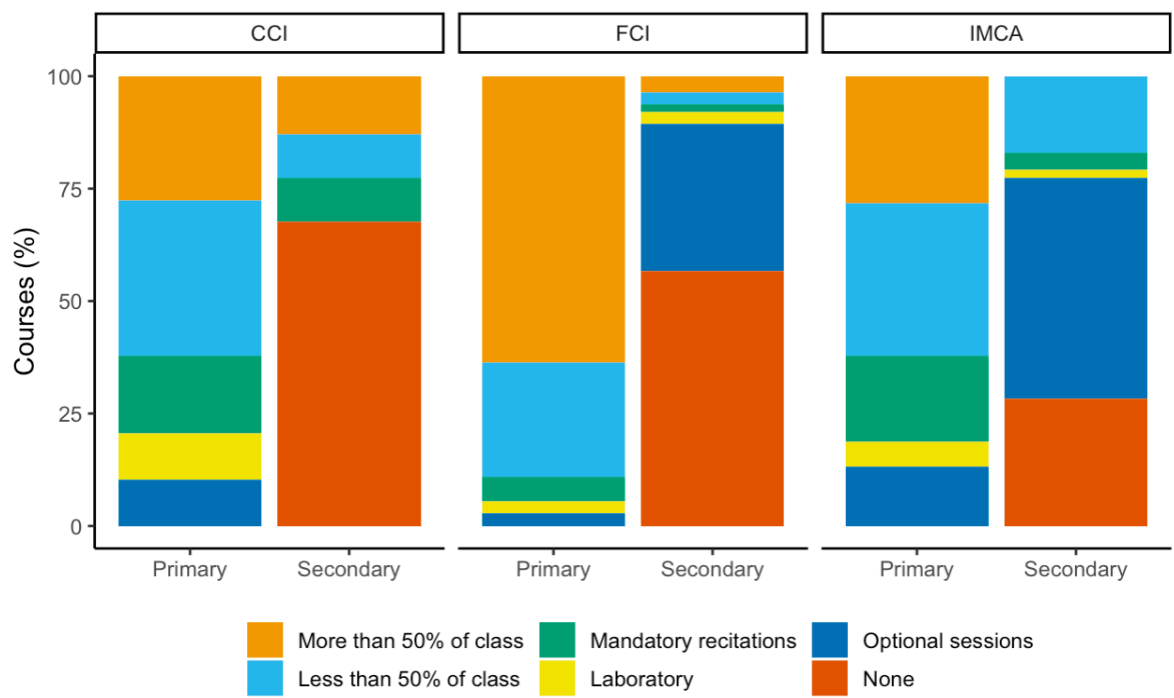

*Appendix Figure 3. Model development workflow with and without priors.*

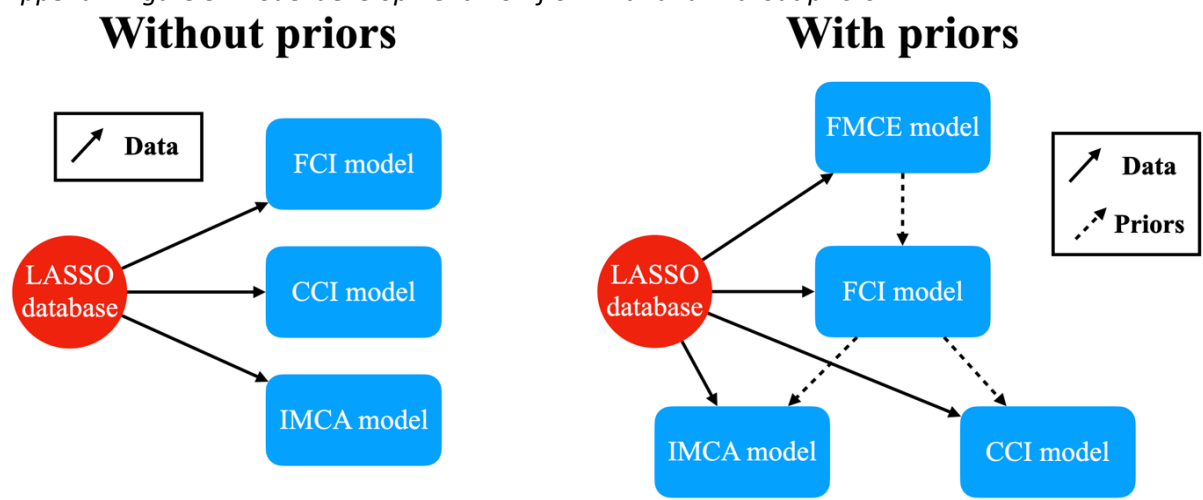

Appendix Figure 4. Violin plots, boxplots, and scatter plots of the data after imputation for pretest and posttest on the (a) Force Concept Inventory (FCI), (b) Introductory Molecular and Cell Biology Assessment (IMCA), and (c) Chemical Concepts Inventory (CCI). The violin plot is a reflected density plot to show the distribution of the data. The notched boxplots show the distribution but focus on the medians with notches to show the 95\% confidence intervals. The scatter plot is jittered to randomly distribute the points left and right of the vertical axis for clarity. The scatter plot illustrates the number of data points in each group and details how the data is distributed, particularly in the tails. The plots show no indication of ceiling or floor effects. These plots show easily identifiable differences across groups but also show that these differences are shifts in the distributions of scores, not gaps that separate groups.

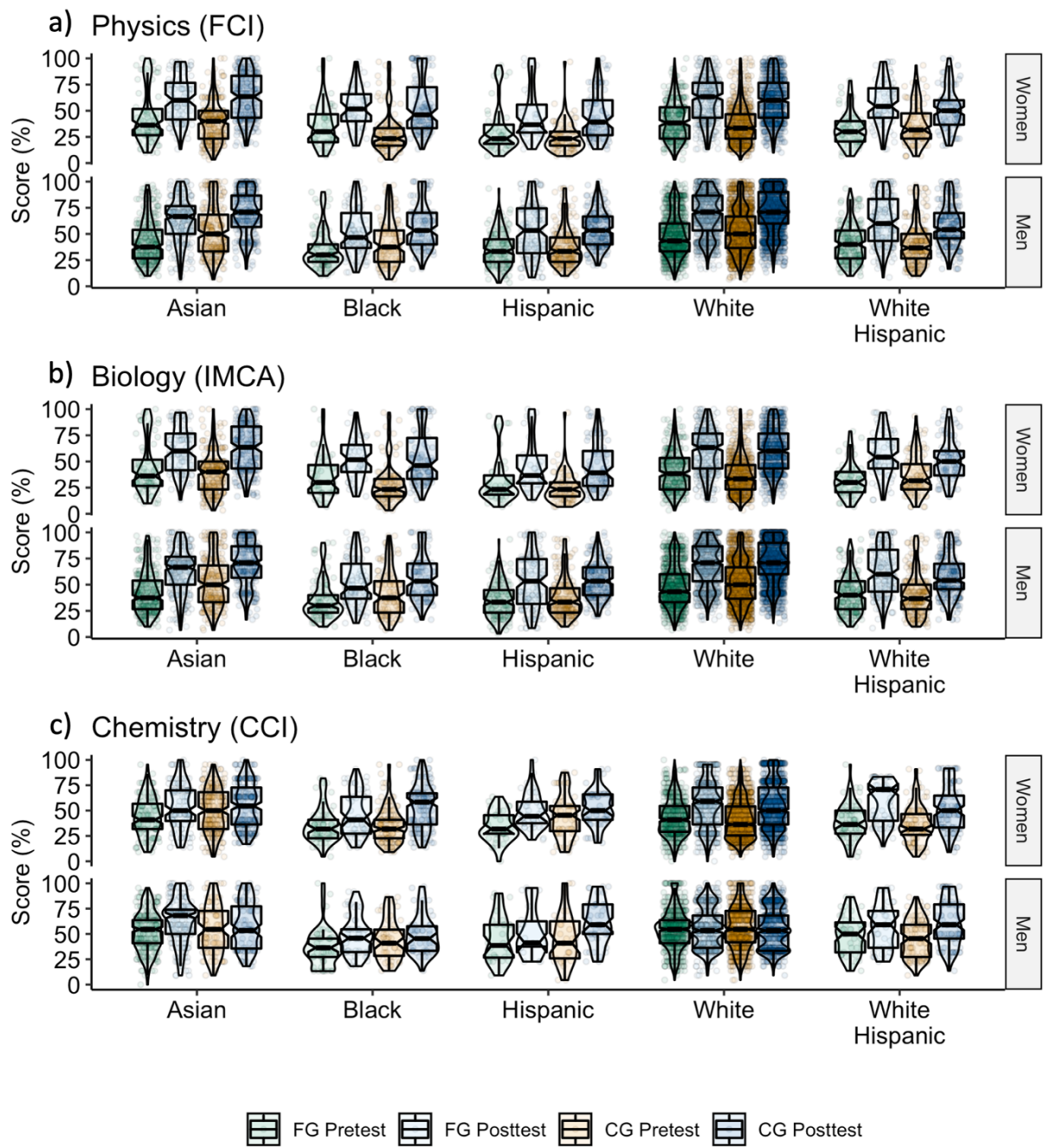